# Efficient online cross-covariance monitoring with incremental SVD: An approach for the detection of emerging dependency patterns in IoT systems

Xinmiao Luan[1], Qing Zou[2], Jian Li[2] and Andi Wang[1*]

*1 The School of Manufacturing Systems and Networks, Arizona State University, Mesa, Arizona, 85212, USA*

*2 Xi'An Jiaotong University, Xi'An, China*

**Abstract**

The development of the manufacturing systems has made it increasingly necessary to monitor the data generated by multiple interconnected subsystems with rapid incoming of samples. Based on incremental Singular Value Decomposition (ISVD), we develop a general online monitoring approach for the relationship of data generated from two interconnected subsystems, where each subsystem produces big data streams with several variation patterns in normal working condition. When special situation happens and new associations occur, a very small amount of computation is sufficient to update the system status and compute the control statistics by using this approach. The proposed method reduces computational overhead and retains only a small number of pairs of possible dependent patterns at each step. The validation of the method through simulation studies and a case study on semiconductor manufacturing processes further supports its effectiveness.

## 1. Introduction

In the past decades, advancements in sensing technology enabled collecting production data from engineering systems to perform data-driven monitoring, modeling, and process adjustment. Complex, interconnected systems are typically equipped with multiple sensors, collecting data that reflect the operation of individual system components. A typical example of such systems is the semiconductor fab, which involves numerous manufacturing systems of deposition, lithography, and etching that collectively fabricate the final integrated-circuit products. The utilization of the data from the interconnected systems started within the operation of individual system. For example, statistical process control has been applied to individual steps of semiconductor manufacturing systems. However, the inter-relationship between the multi-sensor data cannot be revealed.

To improve the data silos, there has been an increasing trend to aggregating and processing data from all systems, known as internet-of-things (IoT). With IoT, one essential capability is to detect identify the emerging correlated patterns of data from multiple systems. As an example, the production engineers in semiconductor manufacturing fab are interested to detect the new correlated patterns between the two data sources. From the film deposition step, the thickness of the deposited layer on the wafer is measured at several points on the wafer. In the overlay step, the alignment error of the projected mask and the previous layer is measured at multiple locations on the wafer, which results in an *overlay map*. With the development of new data-transmission mechanism, the thickness map and the overlay map of the products from the two manufacturing steps can be pooled together, while new monitoring algorithm is needed to perform the monitoring tasks.

The above example is a typical scenario where a process monitoring is required based on the measurements from multiple systems. In this article, we aim to develop a general monitoring approach for the emerging correlated patterns of systems due to an emerging root cause. Specifically, this monitoring approach has two characteristics driven by the practical concerns in the manufacturing industry.

- This paper focus on detecting the emerging patterns of two systems. This is not only because of the analytical simplicity, but also because of an important practical concern: within a multi-system IoT network, pooling all data of multiple stations together is often discouraged given the limitation on data storage and real-time processing; besides, pooling the data from every two systems enables that the related systems be identified directly.
- Our monitoring algorithm aims to be only sensitive to the change in the correlation of the data from two systems, not the changes due to faults within the individual IoT system, since industries like semiconductor manufacturing typically inherit existing monitoring software that detect the faults on the individual systems in the IoT network, before data sharing between systems becomes possible. For this reason, it is desirable to only monitor the correlation of the two datasets from two systems.

Although only the correlation of data from two systems are considered, computational challenge is a major concern in certain applications, due to the potential high dimensional measurements obtained from each system as discussed by Wang and Shi (2021), the hardware constraint on the computational systems for IoT highlighted by Liu (2015), and the rapid rate that the data are generated. In the semiconductor manufacturing scenario described above, both deposition and the lithography steps involve tens to hundreds of measurements, and all monitoring algorithms of the entire fab are deployed on a central server so it very desirable to minimize the computational burden.

Beside the computational challenge, another analytical challenge caused by the high dimensionality of the data is the flexibility on the correlation between the data from the two systems. In engineering practice, however, there are typically limited root causes that affect the system when a new situation occurs to the IoT system, and therefore typically only one correlated pattern emerges from the data of two systems at a time. We will leverage this characteristic in our development of the monitoring approach, to accelerate the detection of the change of the cross-covariance.

In summary, we aim to develop a general monitoring approach for the data from two interconnected systems. We consider the general scenario that two systems generate $p$ and $q$ dimensional feature vectors respectively at each time, and we are interested in detecting the change of cross-covariance of these two data streams due to an emerging root cause that affects two subsystems simultaneously. The method we develop is based on incremental SVD, which only requires performing a relatively small amount of calculation to update the SVD of the estimated cross-covariance matrix after collecting one data subgroup. The benefit is that the algorithm achieves a very small time and space complexity at most linear with $p$ or $q$. For this reason, we name our control chart incremental-SVD (ISVD) chart.

The remaining part of the article is organized as follows. In the next section, we briefly review the monitoring and diagnostics of high-dimensional data, especially the monitoring algorithms based on covariance structure. We will highlight the connections and differences between our approach and these existing methods. In Section 3, we describe our monitoring methodology. Section 4 performs the simulation study to compare the performance our online monitoring algorithm with a computationally-intensive benchmark without online learning. Finally, Section 5 concludes the article.

## 2. Literature Review

Recall that the primary goal of this paper is to detect if there are new correlated variation patterns from the data collected from two systems. There are multiple offline analytics approaches that tackle this problem in literature. For example, Wang et al. (2022) proposed a retrospective analysis approach based on dictionary learning method, which simultaneously identifies the time frame where certain variation patterns of all monitoring signals appear and characterize the variation patterns to reveal the root causes. Another example is Zhang et al. (2020b), which identify the time segments where certain groups of variables subside in linear subspaces, i.e., correlated with each other. However, there are two limitations with this line of research: (1) these methods are not specifically developed for the appearance of correlation from the two systems (2) these methods are offline instead of online, thus not aimed for finding the new correlated patterns with computational limit.

In literature, the online detection of correlated patterns of two datasets is also related to the covariance monitoring problem of the multivariate statistical process control based on Yeh's (2006b) review. However, few studies are based on the assumption that the covariance matrix is subject to a rank-1 change, indicating a single root cause appears to the IoT system in our application. Xie et al. (2020a) developed a sequential monitoring method for detecting the rank-1 change of the covariance matrix. However, they are not specifically aimed for the cross-correlation between two datasets. The existing approaches for modeling the relationship between two datasets include canonical correlation analysis and partial least square approaches, whereas these methods are not suitable for online monitoring.

## 3. Methodology

In this paper, we will develop methodologies for detecting the appearance of correlated patterns in two systems using online learning techniques. In Section 3.1, we will first discuss the formulation of the problem. The incremental SVD method will be introduced in Section 3.2. In Section 3.3, we will discuss the utilization and the extension of the methods.

### *3.1. Problem formulation*

We consider two interconnected systems that generates multivariate data $x \in \mathbb{R}^p$ and $y \in \mathbb{R}^q$ individually at each time. These systems may be consecutive stages in a multistage manufacturing system according to Wang and Shi (2021), or generally two networked modules in an IoT. Under the normal working conditions, we assume that there are $J$ underlying latent factors that drives the of variations of $x$ and $y$ simultaneously, where $J$ is typically a small number because the operations of the two modules are relatively independent. Therefore, we model the observations under normal conditions as

$$x = \mu_x + \sum_{j=1}^{J} z_j u_{0j} + \epsilon_x; \quad y = \mu_y + \sum_{j=1}^{J} z_j v_{0j} + \epsilon_y. \tag{1}$$

In this representation, the latent factor $j$ is represented with a random variable $z_j \sim N(0, s_{0j}^2)$ that cause the pattern $u_{0j}$ and $v_{0j}$ in the data streams $x$ and $y$ respectively, where $s_{0j} > 0, j = 1, \ldots, J$.

The mean of $x$ and $y$ are denoted as $\mu_x$ and $\mu_y$, and they can be estimated from the historical data. Without loss of generality, we assume that they are zeros, as one can always deduct the mean from the samples during the online monitoring process. The sources of variations that affect each individual system is represented as $\epsilon_x$ and $\epsilon_y$. Under this setup, the cross-covariance between $x$ and $y$ is $\Sigma_0 := \text{cov}[x, y] = \sum_{j=1}^{J} s_{0j}^2 \boldsymbol{u}_{0j} \boldsymbol{v}_{0j}^\top$.

According to the assumption of limited root causes, we focus on detecting the appearance of a single root cause that affects both systems through introducing a new latent factor to both $x$ and $y$. After the root cause appears at time $\tau$, the observations are modelled as

$$x = \mu_x + \sum_{j=1}^{J} z_j \boldsymbol{u}_{0j} + z\boldsymbol{u} + \epsilon_x; \quad y = \mu_y + \sum_{j=1}^{J} z_j \boldsymbol{v}_{0j} + z\boldsymbol{v} + \epsilon_y \tag{2}$$

where $z \sim N(0, s^2)$ is the emerging assignable cause. This root cause leads to a new pair of patterns on the data $x, y$, described by vectors $\boldsymbol{u}$ and $\boldsymbol{v}$, where $\boldsymbol{u}, \boldsymbol{v}$ and $s > 0$ are unknown. The post-change cross covariance between $x$ and $y$ becomes $\Sigma = \Sigma_0 + s^2 \boldsymbol{u}\boldsymbol{v}^\top$, characterized by adding a rank-one component of $s^2 \boldsymbol{u}\boldsymbol{v}^\top$ onto $\Sigma_0$.

To detect the change of the system, observations of $(x, y)$ are collected from the process. At each time $t$, a small group of $m$ independent samples $\mathcal{D}_t = \{(x_{t,i}, y_{t,i})\}_{i=1}^{m}$ are collected. When $t < \tau$, the statistical model of $(x_{t,i}, y_{t,i})$ is described by (1), and after $t \geq \tau$, the model of $(x_{t,i}, y_{t,i})$ is described by (2). Based on the sequential measurements, we aim to detect the change point $\tau$ as soon as possible.

The key idea of our proposed scheme is to estimate $s^2$, the largest singular value of $\Sigma - \Sigma_0$ incrementally as the samples $\mathcal{D}_1, \mathcal{D}_2, \ldots$ arrive and use it as a monitoring statistic. The matrix $\Sigma - \Sigma_0$, representing the change of the covariance between $x$ and $y$ after the root cause appears, is rank one.

### 3.2. Cross-covariance monitoring with incremental SVD

To estimate and monitor $s^2$, we start with obtaining estimation of $\Sigma$, the covariance between $x$ and $y$ based on the subgroup of $m$ samples at time $t$. The natural covariance estimation is obtained as $\hat{\Sigma}_t = \frac{1}{m}\sum_{i=1}^m x_{t,i} y_{t,i}^\top$. We apply an EWMA to the sequence $\hat{\Sigma}_t - \Sigma_0$ to refine the estimation with the recent samples, and obtain

$$\mathbf{D}_t = (1-\lambda)\mathbf{D}_{t-1} + \lambda\left(\frac{1}{m}\sum_{i=1}^m x_{t,i} y_{t,i}^\top - \Sigma_0\right). \tag{3}$$

When the process is in control, we shall have $\mathbb{E}[\mathbf{D}_t] = 0$, and when the process change appears, $\mathbb{E}[\mathbf{D}_t]$ will have a rank-one change and become $s^2 uv^\top$. Therefore, the largest singular value of $\mathbf{D}_t$ can be calculated after each time $t$ to obtain the monitoring statistics $s^2$. However, major computational challenge is that $\mathbf{D}_t$ is a big matrix of size $p \times q$, and thereby it is computationally inefficient to compute $\|\mathbf{D}_t\|_*$ at each time. Recall that both $m$ and the rank of $\Sigma_0 = \sum_{j=1}^J s_{0j}^2 u_{0j} v_{0j}^\top$ are small numbers, and therefore (3) can be regarded as a low-rank modification of the matrix $\mathbf{D}_t$. For this reason, we leverage following proposition from the incremental SVD algorithm proposed by Brand (2006a) for the update. The benefit is on the computational efficiency: compared with the full SVD of $\mathbf{X}$ with the time complexity of $O(pq \cdot \min(p,q))$, the incremental SVD result in a complexity of $O(pqr)$, where $r$ is the rank of $\mathbf{X}$.

**Proposition (Brand's Incremental SVD)**: If $\mathbf{X} \in \mathbb{R}^{p \times q}$ and $\mathbf{X}$ has the singular value decomposition $\mathbf{X} = \mathbf{USV}^\top$. The matrix $\mathbf{X} + \mathbf{ab}^\top$ has the singular value decomposition:

$$\mathbf{X} + \mathbf{ab}^\top = \mathbf{U}_1 \mathbf{S}_1 \mathbf{V}_1^\top$$

where $\mathbf{U}_1 = \begin{bmatrix} \mathbf{U} & \frac{\eta}{\|\eta\|} \end{bmatrix} \mathbf{U}_K$; $\mathbf{V}_1 = \begin{bmatrix} \mathbf{V} & \frac{\xi}{\|\xi\|} \end{bmatrix} \mathbf{V}_K$, $\boldsymbol{\eta} = (\mathbf{I} - \mathbf{UU}^\top)\mathbf{a}$ is component of $\mathbf{a}$ orthogonal to the column space of $\mathbf{U}$, and $\boldsymbol{\xi} = (\mathbf{I} - \mathbf{VV}^\top)\mathbf{b}$ is the component of $\mathbf{b}$ orthogonal to $\mathbf{V}$, and $\mathbf{K} = \mathbf{U}_K \mathbf{S}_1 \mathbf{V}_K^\top$ gives the SVD of $\mathbf{K} = \begin{bmatrix} \mathbf{S} & 0 \\ 0 & 0 \end{bmatrix} + \begin{bmatrix} \mathbf{U}^\top \mathbf{a} \\ \|\eta\| \end{bmatrix} \begin{bmatrix} \mathbf{b}^\top \mathbf{V} & \|\xi\| \end{bmatrix}$.

Using the Incremental SVD, we can update the SVD of $\mathbf{D}_t = \mathbf{U}_t \mathbf{S}_t \mathbf{V}_t^\top$ through $m + J$ increments. Besides the application of Brand's Algorithm, we propose to threshold the singular values of $\mathbf{D}_t$ after each step, by only keep the leading $r$ components of the SVD, where $r \in \mathbb{N}_+$ is a tuning parameter. The purpose is to further reduce the time and space complexity required by the Brand algorithm during the monitoring process, and at the same time avoid the curse of dimensionality. This measure can be effective even if $r$ is selected not too big, because the rank of $\mathbb{E}[\mathbf{D}_t]$ is at most 1. The algorithm is summarized in Algorithm 1.

---

**Algorithm 1: online emerging association monitoring based on incremental SVD**

Inputs: The SVD of $\Sigma_0$, $\{\mathbf{u}_{0j}, \mathbf{v}_{0j}, s_{0j}\}_{j=1}^J$, obtained from the retrospective analysis of historical in-control data; the prescribed EWMA coefficient $\lambda$; the maximal rank of $\mathbf{D}_t$, $r$; the control limit $H$.

1: Initiate $\mathbf{S}_0 = \mathbb{R}^{0 \times 0}$, $\mathbf{U}_0 \in \mathbb{R}^{p \times 0}$, $\mathbf{V}_0 \in \mathbb{R}^{q \times 0}$ as empty arrays.
2: For $t = 1, 2, \ldots$
3:     $\mathbf{S}_t \leftarrow m \frac{1-\lambda}{\lambda} \mathbf{S}_{t-1}$
4:     For $i = 1, \ldots, m$:
5:         $[\mathbf{U}_t, \mathbf{S}_t, \mathbf{V}_t]$ = updateSVD$(\mathbf{U}_{t-1}, \mathbf{S}_t, \mathbf{V}_{t-1}, \mathbf{x}_{t,i}, \mathbf{y}_{t,i})$
6:     $\mathbf{S}_t \leftarrow \mathbf{S}_t / m$
7:     For $j = 1, \ldots, J$:
8:         $[\mathbf{U}_t, \mathbf{S}_t, \mathbf{V}_t]$ = updateSVD$(\mathbf{U}_t, \mathbf{S}_t, \mathbf{V}_t, s_{0j} \mathbf{u}_{0j}, -s_{0j} \mathbf{v}_{0j})$
9:     $\mathbf{S}_t \leftarrow \lambda \mathbf{S}_t[1:r, 1:r]$; $\mathbf{U}_t \leftarrow \mathbf{U}_t[:, 1:r]$; $\mathbf{V}_t \leftarrow \mathbf{V}_t[:, 1:r]$

| 10: | Monitoring Statistic $T_t = \mathbf{S}_t[1,1]$ |
| 11: | Trigger alarm if $T_t > H$. |

In Algorithm 1, Line 3-9 update the SVD of $\mathbf{D}_t$ according to Equation (3), which can be shown by induction. Given the SVD of $\mathbf{D}_{t-1} = \mathbf{U}_{t-1}\mathbf{S}_{t-1}\mathbf{V}_{t-1}^\top$, Line 3 calculates SVD $\frac{m(1-\lambda)}{\lambda}\mathbf{D}_{t-1} = \mathbf{U}_{t-1}\left(\frac{m(1-\lambda)}{\lambda}\mathbf{S}_{t-1}\right)\mathbf{V}_{t-1}^\top$. The "for" loop from Line 4-5 further gives the SVD of $\frac{m(1-\lambda)}{\lambda}\mathbf{D}_{t-1} + \sum_{i=1}^m \mathbf{x}_{t,i}\mathbf{y}_{t,i}^\top = \mathbf{U}_t\mathbf{S}_t\mathbf{V}_t^\top$, and Line 6 updates the SVD to match $\left[\frac{m(1-\lambda)}{\lambda}\mathbf{D}_{t-1} + \sum_{i=1}^m \mathbf{x}_{t,i}\mathbf{y}_{t,i}^\top\right]/m = \mathbf{U}_t\mathbf{S}_t\mathbf{V}_t^\top$. Finally, Line 7-8 updates the SVD to match $\left[\frac{m(1-\lambda)}{\lambda}\mathbf{D}_{t-1} + \sum_{i=1}^m \mathbf{x}_{t,i}\mathbf{y}_{t,i}^\top\right]/m - \sum_{j=1}^J s_{0j}^2 \mathbf{u}_{0j}\mathbf{v}_{0j} = \mathbf{U}_t\mathbf{S}_t\mathbf{V}_t^\top$, which lead to

$$\mathbf{D}_t = (1-\lambda)\mathbf{D}_{t-1} + \lambda\left[\frac{1}{m}\sum_{i=1}^m \mathbf{x}_{t,i}\mathbf{y}_{t,i}^\top - \Sigma_0\right] = \mathbf{U}_t[\lambda\mathbf{S}_t]\mathbf{V}_t.$$

Finally, line 9 of the algorithm approximates $\mathbf{D}_t$ with a rank-$r$ matrix by only keeping the first $r$ components in the SVD.

### 3.3. Implementation of the control chart

The Algorithm 1 provides an overall procedure for updating the charting statistics with the new sample $\mathcal{D}_t$. When implementing this control chart, we first estimate the in-control cross-covariance $\Sigma_0 = \sum_{j=1}^J s_{0j}^2 \mathbf{u}_{0j}\mathbf{v}_{0j}$ using sample cross-covariance of the in-control historical datasets. After we select the value of tuning parameters $\lambda > 0$ and $r \in \mathbb{N}_+$, we can use the in-control historical data to search the control limit $H$ that gives a specific value of the in-control ARL. The algorithm is ready to be used for monitoring the online data streams $\{\mathcal{D}_t\}$.

### *3.4. Summary and Discussions*

In summary, the ISVD chart introduced above integrates three ideas:

(1) the EWMA control chart that gives the new samples with higher weights to estimate the recent cross-covariance matrix.

(2) Thresholding $\mathbf{D}_t$, the estimated modification of the cross-covariance matrix at rank $r$, to eliminate the noise or less significant relationships from the data streams, enhance the important relationships between the data from two systems, and to alleviate the computational burden.

(3) Using Brand's incremental SVD Algorithm for further alleviating the computational burden.

For ISVD algorithm, each time involves $J + m$ incremental SVD operations. Therefore, the total computational complexity of each time step is $O(pqr(J + m))$. When $r(J + m) \ll \min(p, q)$, the utilization of the ISVD method would significantly improve the computation efficiency. As will be detailed in simulation studies, we can see that the effectiveness of the ISVD chart is not sensitive to $r$, and thereby a small value is preferred. Besides, the ISVD method is especially efficient when the in-control data from two systems have limited number correlated patterns ($J$ is small) and the subgroup is small ($m$ is small).

We also note that the ISVD chart can be extended to special data types generated from the IoT systems, although they are based on a simple situation where the data from two systems $x$ and $y$ are general data vectors. If the data generated from the two systems have more complex structures, such as signals and images in Wang and Shi's study (2021), one possible approach is to extract a comprehensive set of features from the data from each system that determines its status. Such features can be either based on engineering knowledge or using data-driven feature extraction and dimension reduction techniques.

## 4. Simulation studies

In this Section, we conduct simulation studies to evaluate the performance of the proposed online monitoring method. We noted that there is no existing approach of monitoring the cross covariance. Therefore, we focus on comparing our ISVD chart with the benchmark scheme of monitoring the largest eigen vector of $\hat{\Sigma}_t - \Sigma_0$. For a fair comparison, we perform EWMA of $\hat{\Sigma}_t - \Sigma_0$, and thereby the benchmark scheme monitors the largest singular value of $\mathbf{D}_t$ in Equation (3). Compared with ISVD approach, the benchmark approach (1) does not use the incremental SVD to alleviate the computational burden and (2) does not apply the thresholding of $r$.

### 4.1. Simulation Setup

The goal of the simulation study is to evaluate the performance of the proposed method when the data's characteristics and the configuration of the proposed methods vary and compare the proposed method with the benchmark method. We consider a total of 9 simulation setups. In the first setup, we consider the scenario where $X$ and $Y$ are independent when the process is in-control and thereby $J = 0$, and the chart is designed with $\lambda = 0.02$ and $r = 5$. When the process becomes out-of-control, a new pair of variation patterns $\mathbf{u}, \mathbf{v}$ appear according to Equation (2) and the magnitude is represented as $s^2 = 0.5, 1,$ and 2. In the other setups, we vary the value of $\lambda, r$ and the in-control and out-of-control data. The setups are listed in Table 1:

1. Setup 2-3 evaluates the effect of parameter $\lambda$ on the chart performance, where $\lambda$ is changed to 0.01 and 0.05 respectively.

2. Setup 4-5 evaluates the effect of parameter $r$ on the chart performance, where $r$ is changed to 2 and 10.

3. Setup 6-9 evaluates how the chart performs when the in-control $X$ and $Y$ are correlated ($J = 1$). Specifically, setup 6-7 and 8-9 differs in the specifications of the out-of-control scenario:

   a. In setup 6-7, the new variation patterns $u, v$ are *parallel* to the existing correlated patterns $u_{01}, v_{01}$ respectively.

   b. In setup 8-9, the new variation patterns $u, v$ are *perpendicular* to in-control pattern $u_{01}, v_{01}$.

   Both the setups 6-7 and the setups 8-9 are different in the magnitude of the existing correlations, described by the parameter $s_{01}$.

Apart from the above specific setups, below are common parameters of all 9 setups: the dimension of $x$ and $y$ are 10 and 20 respectively, the sample size $m = 5$, and $\epsilon_x \sim N(0, I), \epsilon_y \sim N(0, I)$. The value of $u_{01}, v_{01}$ are generated from uniform distribution on sphere.

Table 1 The simulation setups

|  | Interpretation | $J$ | $s_{01}$ | $u, v$ | Proposed $\lambda$ | $r$ |
|---|---|---|---|---|---|---|
| Setup 1 | Baseline | 0 | NA | NA | 0.02 | 5 |
| Setup 2 | Small $\lambda$ | 0 | NA | NA | 0.01 | 5 |
| Setup 3 | Big $\lambda$ |  |  |  | 0.05 |  |
| Setup 4 | Small r | 0 | NA | NA | 0.02 | 2 |
| Setup 5 | Big r |  |  |  |  | 10 |
| Setup 6 | Small IC correlation, parallel OC patterns | 1 | 0.5 | $\parallel u_{01}, v_{01}$ | 0.02 | 5 |
| Setup 7 | Big IC correlation, parallel OC patterns |  | 1 |  |  |  |
| Setup 8 | Small IC correlation, perpendicular OC patterns | 1 | 0.5 | $\perp u_{01}, v_{01}$ |  |  |
| Setup 9 | Big IC correlation, perpendicular OC patterns |  | 1 |  |  |  |

The control limits of each chart are set respectively so that an in-control ARL of 1000 is achieved.

## 4.2. Simulation Results

The simulation results are demonstrated in Figure 1. The messages from the simulation studies and the interpretations of the subfigures are as follows:

1. **The selection of $\lambda$ in ISVD should be comparable with the size of the shift.** The first figure evaluates the effect of $\lambda$ on the ARL of the charts under the out-of-control conditions based on simulation setups 1-3 and compare them with the benchmark method with various values of $\lambda$. When the change size is small ($s^2 = 0.5$), we find that $\lambda = 0.02$ gives the best chart performance as the ARL of the solid red curves and the dashed red curves are the smallest among all solid and dashed curves. When the change size is bigger ($s^2 = 1$ or 2), the blue lines ($\lambda = 0.05$) achieve the minimal out-of-control ARL. It confirms with the general rule for EWMA chart that larger $\lambda$'s should be applied if bigger sizes of changes are expected. Besides, we see that the observed method outperforms the benchmark method.

2. **Smaller values of $r$ is preferred as only one new pattern appears.** The second subfigure describes the effect of $r$ on the performance of the proposed chart and compare it with the benchmark monitoring scheme. From this figure, we can see that small value of $r$ is preferred, as there is only one pair of new patterns $\boldsymbol{u}$ and $\boldsymbol{v}$ that present in the data of $\boldsymbol{x}$ and $\boldsymbol{y}$. The proposed method always outperforms the benchmark. Essentially, this can be interpreted because when $r = \min(p, q)$, the proposed method and the benchmark method will be the same analytically, despite that their procedures of calculating the SVD are different.

3. **Given dependent $x$ and $y$, the ISVD chart is more effective if the new pair of emerging new patterns are perpendicular to the in-control correlated patterns.** The third and the fourth subfigure describes the effect of in-control correlation, under the situations that the new pair of patterns $u$ and $v$ under the out-of-control states are either parallel or perpendicular to the existing patterns. When the new patterns are parallel to the existing patterns, we can see that larger existing correlation will cause the chart less sensitive to shifts. However, this effect is not significant when the new patterns are perpendicular with the existing patterns, and the performance of the chart is significantly better as the ARL becomes smaller with the same size of shift. In all cases, the proposed chart outperforms the benchmark.

Apart from the information from the simulation results, it is worth to mention that the computational complexity of the proposed method is smaller than the proposed approach given that the values of $J$ and $m$ are not too big, which can be validated through the analysis of the algorithms' complexities in Section 3.

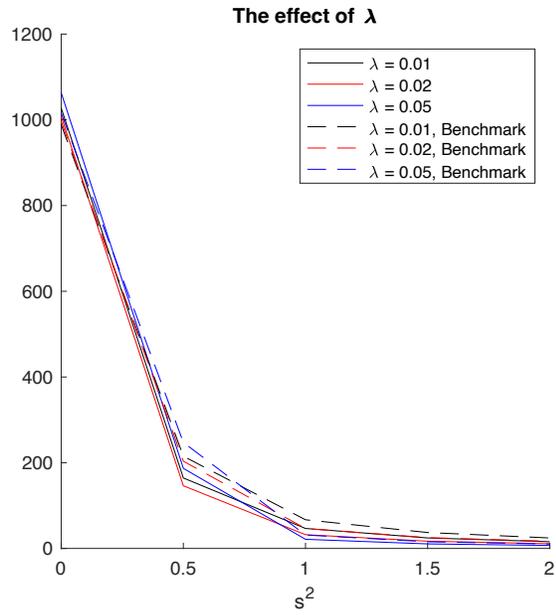
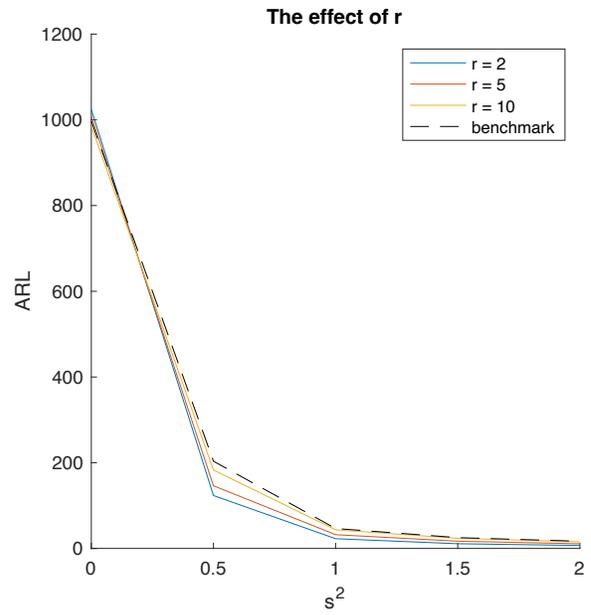
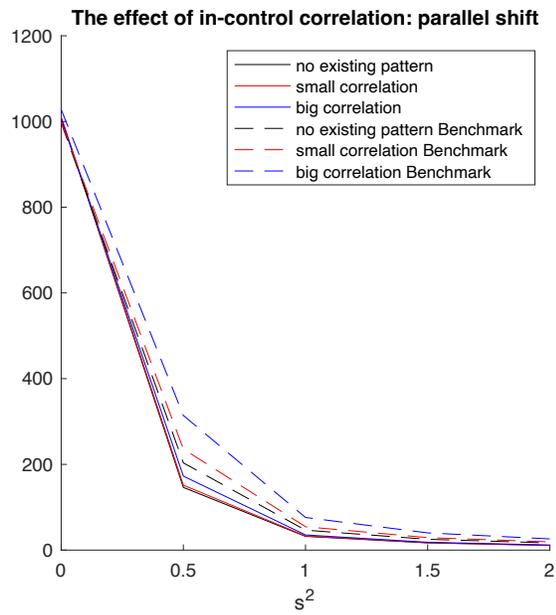
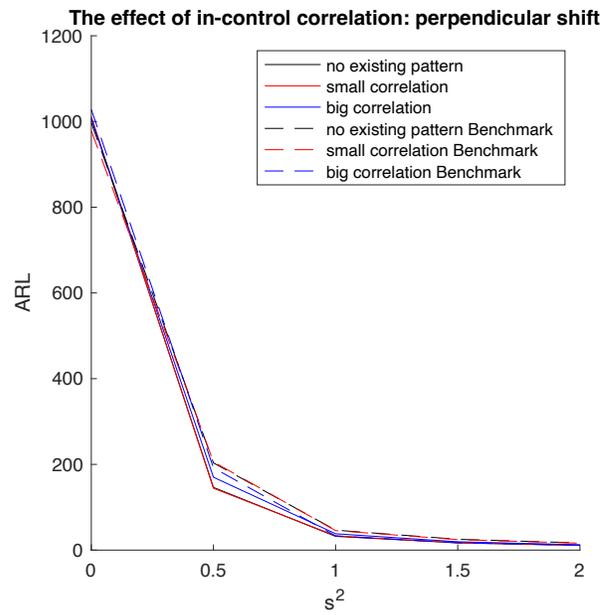

Figure 1 Simulation Results

## 5. Case study

In this section, we demonstrate the proposed online monitoring method with an example of semiconductor manufacturing.

In semiconductor manufacturing process, lithography and deposition are two critical manufacturing steps. The lithography process prints the project mask of designed circuit patterns onto the wafer surface. Therefore, the misalignment between the actual printed location on the wafer surface and its designed location is a critical quality measurement. In practice, the overlay error is measured at several locations on the wafer, determined by the product design. The misalignment error at each location can be represented as a 2D vector. These measurements are shown in the left panel of Figure 2 as an overlay map, where the $X_1 X_2$ plane represents the wafer surface, and each vector indicates the overlay error measured at that location. The overlay error map is affected by many types of geometric errors and ideally, an overlay error map should exhibit short and random vectors, according to Armitage et al. (1988) and Huang et al. (2008). The overlay error measurements can be represented with $p$ variables, where $p$ is two times the sites of the measurement locations. In the deposition process, a thin layer of material is deposited onto the wafer surface. After the deposition process, the thickness of the deposited layer is measured at $q$ locations on the wafer, and therefore the thickness map can be represented as $q$ variables. The deviation of the thickness from its nominal value is shown in the right panel of Figure 2.

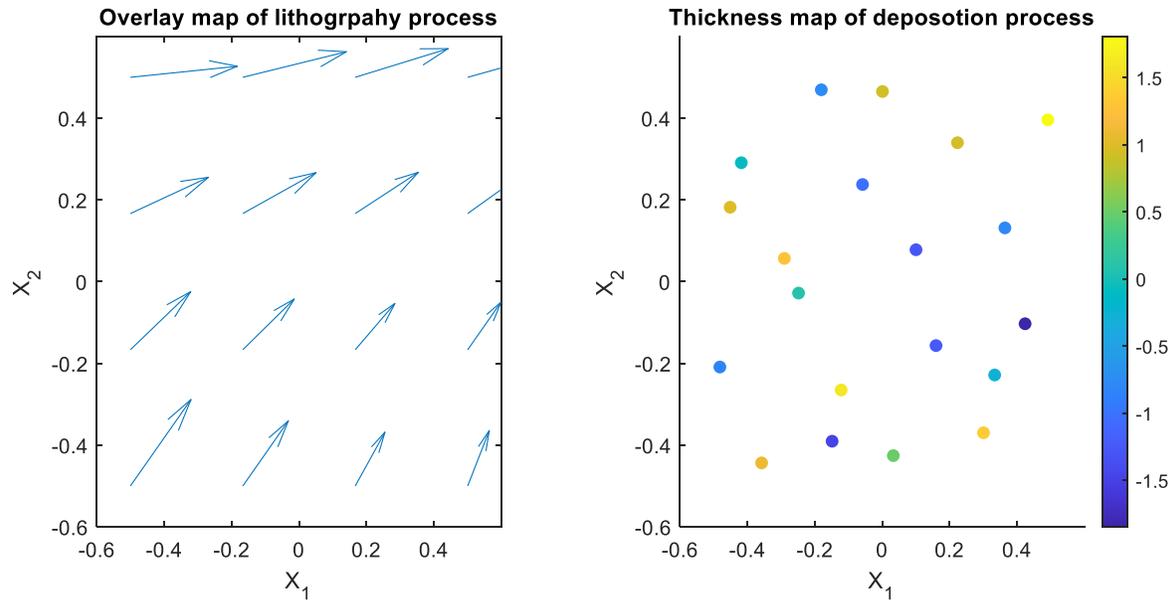

Figure 2 The overlay map and the thickness map of lithography and overlay steps of semiconductor manufacturing.

Previously, the data from deposition and lithography steps in a manufacturing process are analyzed individually for the improvement of individual steps. After the data from both steps can be aggregated, the semiconductor manufacturing engineers are interested in identifying the relationship between the overlay map of the lithography step and the thickness map of the deposition step. In this case study, we test our approach with a *simulation test bed* where the overlay and deposition data of consecutive wafers are generated. This simulation testbed is configured so that the overlay pattern of $x$-axis translational error becomes dependent with the wafers' slope on a specific direction starting after sample $\tau = 100$. The simulation test bed is necessary as it enables us to know the precise time of the change point while the characteristics of the real production data is preserved.

The control chart is illustrated in Figure 3. The horizontal line represents the control limit $H = 3.6$ that achieves an in-control ARL = 600, and the charting statistics exceed this control limit at time 124, with a detection delay of 24. It illustrates the effectiveness of the method.

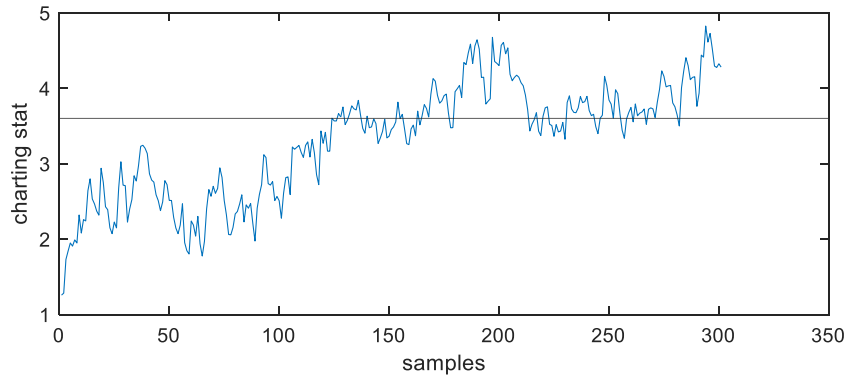

Figure 3 The control chart of the semiconductor manufacturing case study.

It is worth mentioning that in this case study, the specific locations of the thickness map and the overlay map are ignored. In future study, we will investigate how to incorporate the spatial information into our ISVD monitoring approach.

## 6. Conclusion

In this paper, we proposed an incremental SVD approach that performs online cross-covariance monitoring. This approach can be used for detecting the emerging dependent variation patterns from two systems in IoT. Our proposed method is based on incremental SVD, which significantly helps to reduce the computational overload. Besides, it keeps only a small number of pairs of possible dependent patterns in every step. The proposed method is validated with both simulation study and the data generated from the thickness and overlay data of semiconductor manufacturing processes.

## Data Availability Statement

The data that support the findings of this study are available from the corresponding author upon reasonable request.